\documentclass[cmp]{svjour}
\usepackage{times}
\usepackage{amsmath}
\usepackage{amsfonts,amssymb}

\journalname{Communications in Mathematical Physics}

\def\be{\begin{equation}}
\def\ee{\end{equation}}
\def\ba{\begin{eqnarray}}
\def\ea{\end{eqnarray}}
\newcommand{\er}{\eqref}

\def\Mapsto{\longmapsto}
\def\To{\longrightarrow}
\def\V{\mathcal{V}}
\def\M{\mathcal{M}}
\def\H{\mathcal{H}}
\def\B{\mathcal{B}}
\def\D{\mathcal{D}}
\def\L{\mathcal{L}}

\def\T{\mathcal{T}}
\def\calC{\mathcal{C}}

\def\C{\mathbb{C}}
\def\Z{\mathbb{Z}}
\def\R{\mathbb{R}}
\def\N{\mathbb{N}}
\def\Q{\mathbb{Q}}

\def\d{\partial}
\def\tri{\triangle}

\def\eps{\varepsilon}
\def\vp{\varphi}
\def\ind{\operatorname{ind}}
\def\rank{\operatorname{rank}}

\def\Tr{\operatorname{Tr}}

\def\tr{\operatorname{tr}}
\def\span{\operatorname{Span}}
\def\<{\langle}
\def\>{\rangle}
\def\half{\frac{1}{2}}

\def\zbar{\bar{z}}

\def\jbar{\bar{j}}

\newcommand{\dslash}{{\d \hspace{-6pt} \slash}}

\def\dpl{\d_+}
\def\dm{\d_-}

\def\dpl{\d_+}
\def\dm{\d_-}

\def\Ind{\operatorname{Ind}}
\def\Ker{\operatorname{Ker}}
\def\Coker{\operatorname{Coker}}
\def\uInd{\underline{\Ind}}
\def\uKer{\underline{\Ker}}
\def\uCoker{\underline{\Coker}}

\def\prooflem#1{\vskip 0.1 in \noindent {\it Proof of Lemma \ref{#1}. } $\ $ }
\def\prooftheorem#1{\vskip 0.1 in \noindent {\it Proof of Theorem \ref{#1}. } $\ $ }
\def\endproof{ $\ \ \Box$ \vskip 0.1 in }

\def\tilde#1{\widetilde{#1}}
\def\norm#1{\left\| #1 \right\|}
\def\abs#1{\left| #1 \right|}
\def\ket#1{|\, #1 \,\rangle}
\def\bra#1{\langle \, #1 \, |}
\def\braket#1#2{\langle \, #1 \, | \, #2 \, \rangle}
\def\t#1{\tilde{#1}}
\def\b#1{\mathbf{#1}}

\def\at#1#2{ \left. #1 \right|_{#2}}
\def\te#1{\text{#1}}
\def\bar#1{{\overline{#1}}}
\def\ex#1{\left\langle #1 \right\rangle}
\def\f#1#2{\frac{#1}{#2}}

\def\JW{{\cal J}\!{\cal W}}

\begin{document}

\title{{Vacuum Geometry of the $N=2$ Wess-Zumino Model}}
\titlerunning{{Vacuum Geometry of the $N=2$ Wess-Zumino Model}}

\author{William Gordon Ritter}
\authorrunning{Gordon Ritter}

\institute{Harvard University Department of Physics, 17 Oxford
St., Cambridge, MA 02138}

\communicated{J. Imbrie}

\maketitle

\begin{abstract} We give a mathematically rigorous construction of
the moduli space and vacuum geometry of a class of quantum field
theories which are $N=2$ supersymmetric Wess-Zumino models on a
cylinder. These theories have been proven to exist in the sense of
constructive quantum field theory, and they also satisfy the
assumptions used by Vafa and Cecotti in their study of the
geometry of ground states. Since its inception, the Vafa-Cecotti
theory of topological-antitopological fusion, or $tt^*$ geometry,
has proven to be a powerful tool for calculations of exact quantum
string amplitudes. However, $tt^*$ geometry postulates the
existence of certain vector bundles and holomorphic sections built
from the ground states. Our purpose in the present article is to
give a mathematical proof that this postulate is valid within the
context of the two-dimensional $N=2$ supersymmetric Wess-Zumino
models. We also give a simpler proof in the case of holomorphic
quantum mechanics.
\end{abstract}


\section{Introduction}
\label{sec:intro}

The purpose of this paper is to provide a mathematically rigorous
version of the physical theory of $tt^*$ geometry, valid within
constructive $N=2$ quantum field theories.

In the setting of topological string theory, Witten
\cite{Witten:1993ed} has shown that a partial understanding of
background independence may be obtained from the geometry of
theory space. $tt^*$ geometry \cite{Cecotti:1991me} is the theory
of bundles, metrics, connections, and curvature over theory space.
For Calabi-Yau spaces, this subject was studied by Strominger
\cite{Strominger:pd} and by Greene {\it et al.}
\cite{Greene:1993vm} in the context of \emph{special geometry},
which refers to the target-space geometry of $N=2$ supersymmetric
vector multiplets, possibly coupled to supergravity. Moreover, the
seminal work \cite{Bershadsky:1993cx} shows the importance of
$tt^*$ geometry as a powerful tool for calculations of exact
quantum string amplitudes.

The ground state metric, originally introduced as a generalization
of special geometry which is valid off-criticality in RG space, is
a Hermitian metric on a complex vector bundle. The base space of
this bundle is formed from suitable collections of coupling
constants for the theory, while the fiber over a point in moduli
space is built from the ground states of the associated quantum
field theory. For the supersymmetric theories we study, the fibers
may also be described as BRST cohomology of the supercharge
operator.

One goal of this paper is to provide detailed descriptions of the
coupling constant spaces relevant to the $N=2$ Wess-Zumino model
on a cylinder. Verification of the vector bundle axioms in these
models is a quantum field theory version of the problem of
continuity in $\kappa$ of the Schr\"odinger operator $-\Delta +
\kappa V$, thus existence of the vacuum bundle as a vector bundle
in the rigorous sense requires analytic control over operator
estimates. We also discuss the mathematical prerequisites
necessary to define the metric and connection of $tt^*$ geometry.
A further mathematical question is the existence of a special
gauge in which the anti-holomorphic components of the connection
vanish.

\subsection{Constructive Quantum Field Theory}

We work with a class of quantum field theories which are
two-dimensional Euclidean $N=2$ Wess-Zumino models, making some
technical assumptions which ensure that cluster expansion methods
are valid. These interactions are also frequently called
``Landau-Ginzburg'' as the simplest bosonic self-interaction
occurs in Landau and Ginzburg's study of condensed matter. For
classification of $N=2$ theories, and the structure of the closely
associated topological theories, see
\cite{Vafa:1988uu,Lerche:1989uy}. The relation to string theory is
described in \cite{Dijkgraaf:1990dj}.

We begin by defining the theory and recalling some known results.
This is a theory of one complex scalar field $\phi$ and one
complex Dirac fermion $\psi$. The formal Hamiltonian is given by
\be \label{wzham}
    H = H_0 + \int \Bigg( \abs{W'_\lambda(\phi)}^2 - \abs{\phi}^2 +
    \bar{\psi} \Bigg[ \begin{matrix} W_\lambda''(\phi) - 1 & 0 \\ 0 &
    W_\lambda''(\phi)^* - 1 \end{matrix} \Bigg] \psi \Bigg) dx
\ee
where $H_0$ is the free Hamiltonian for a boson and fermion with
unit mass and $W_\lambda(x) \equiv \lambda^{-2} \t{W}(\lambda x)$.
The cluster expansion is known to converge under the following
conditions:
\begin{enumerate}
    \item[(A)] $\t{W}'$ must have $n-1$ distinct zeros
    $\xi_1, \ldots, \xi_{n-1}$, where $n = \deg\t{W}$, and

    \item[(B)] $|\t{W}''(\xi_i)| = 1$ for all $i = 1, \ldots, n-1$.
\end{enumerate}
The bosonic potential $\abs{W_\lambda'(\phi)}^2$ has minima at the
zeros of $\tilde{W}'$ and scaling $\lambda \to 0$ increases the
depth and the separation of the potential wells. Thus for
sufficiently small $\lambda$, semiclassical analysis is valid.

The technical restrictions on our class of superpotentials make
the theory amenable to cluster expansion methods, which have led
to proofs of the existence of the infinite volume limit
\cite{janowskythesis}, and a vanishing theorem \cite{JW,Borgs:gq}
for fermionic zero modes in the finite volume theory.

No new phenomena are expected for sufficiently small perturbations
of the mass=1 condition (B). Moreover, for the present study, this
condition must be removed; with condition (B) in place, the space
of admissible potentials is not an open subset of the natural
Euclidean space into which it is embedded.

As observed by Janowsky {\it et al.} \cite{JW}, the relevant
cluster expansions all continue to hold unchanged for small
polynomial perturbations
\be \label{eq:perturb-poly}
    W_\lambda(z)
    \To
    \lambda^{-2} \tilde{W}(\lambda z) + \lambda^{-1} \beta w(\lambda z)
\ee
where $w \in \C[z]$ is a polynomial of degree $n$ and $\beta$ is a
small parameter. This breaks any artificial symmetry due to the
mass restriction, and enlarges the space of admissible
superpotentials. Our approach to eliminating the mass restriction
is to analyze this symmetry breaking in detail, and to show that
the symmetry breaking perturbation \eqref{eq:perturb-poly} results
in the replacement of the closed condition (B) with a condition
that each $\t{W}''(\xi_i)$ must lie in an appropriately small open
neighborhood of the unit circle. See Sec.~\ref{sec:modulispace}
and in particular, Theorem \ref{thm:fine}.

Imbrie {\it et al.} \cite{Imbrie:1989fq} studied the cluster
expansion for the Dirac operator $i\dslash + m(x)$ with a
space-dependent mass, which is a toy model for the infinite volume
multiphase $N=2$ Wess-Zumino$_2$ theory. The appendix to
\cite{Imbrie:1989fq} gives a method for removal of the $|m(x)| =
1$ restriction for the space-dependent Dirac operator, which
generalizes to the full Wess-Zumino model, giving a second method
for removal of (B).

Integrating out the fermions gives the formal partition function
\be \label{formalZ}
    Z = \int d\mu(\phi) e^{-\int \left( \abs{W'_\lambda(\phi)}^2 -
    \abs{\phi}^2 \right) dx } \det [1 + S \gamma_0
    \chi_\Lambda(Y(W_\lambda''(\phi)) - 1)],
\ee
with
\[
    Y(z) \equiv \Big( \begin{matrix} z & 0 \\ 0 & z^*
    \end{matrix} \Big)
\]
where $d\mu(\phi)$ is the normalized Gaussian measure with
covariance $(-\Delta +1)^{-1}$. Also, $S$ is a fermionic
propagator defined as $S = \gamma_0 (i \dslash + 1)^{-1}$ where
$\dslash = \gamma_\mu^E \d_\mu$, and $\gamma^E$ are Euclidean
gamma matrices defined by
\[
    -i \gamma_0 = \gamma_0^E = \Big( \begin{matrix} 0 & -1 \\ 1 & 0
    \end{matrix} \Big),
    \qquad
    \gamma_1^E = \Big(\begin{matrix} 0 & i \\ i & 0\end{matrix} \Big)
\]

The formal expression (\ref{formalZ}) is not well-defined without
normal ordering. The normal-ordered partition function is
\[
Z = \int d\mu(\phi) e^{- \int_\Lambda \left( :
\abs{W_\lambda'(\phi)}^2 : \,-\, :\abs{\phi}^2: \right) dx } {\rm
det}_3 [1+K(\phi)] e^{-R}
\]
where $K(\phi) = S \gamma_0 \chi_\Lambda (Y(W_\lambda''(\phi)) -
1)$ and $R$ is a counter-term given by
\[
    R \equiv \int_\Lambda dx \left[
        \abs{W_\lambda'(\phi)}^2
        - :\abs{W_\lambda'(\phi)}^2:
        - \abs{\phi}^2 + :\abs{\phi}^2:
    \right]
    + \half \Tr(K^2(\phi))
    - \Tr K(\phi)
\]
Supersymmetry of the theory implies that the counter-term $R$ is
\emph{finite}, which means that if we regularize $R$ then the
limit as the regularization is removed is well-defined. In finite
volume this theory was constructed in \cite{2DWZW} and \cite{JL}
with no restriction on the superpotential. The infinite volume
limit is treated via cluster expansions in \cite{janowskythesis}
and \cite{Janowsky:1990jm}.

\subsection{Supersymmetric Lagrangians}

The transformation properties of the Wess-Zumino model under
supersymmetry become especially transparent when it is written in
terms of a manifestly supersymmetric action, \be
\label{Lagrangian}
    S = \int d^4\theta K(\Phi, \bar{\Phi}) + \int d^2\theta^+ W(\Phi)
    + \int d^2 \theta^- \bar{W}(\bar{\Phi})
\ee
where $\Phi$ is a superfield.

Typically in constructive field theory one restricts attention to
the K\"ahler form
\[
    (- \frac{1}{4} \Phi^* \Phi)
\]
arising from a flat metric, since some work is required to
generalize the cluster expansion to more general $K$. We hope to
address this question in a separate paper, but for the present we
also use the flat K\"ahler form.

Expanding the superfields in lightcone coordinates and eliminating
auxiliary fields from the Lagrangian density \eqref{Lagrangian}
using their equations of motion, one obtains
\begin{eqnarray}
    {\cal L} &=&
    \sum_{i=1}^n \Big(
        \f12 \dpl\vp_i^* \dm\vp_i
        + \f12 \dm\vp_i^* \dpl\vp_i
        - \abs{\d_i W(\vp)}^2
        + i\psi_{1,i}^* \dm\psi_{1, i}
        \nonumber \\
    && \qquad \qquad \qquad\phantom{1}
        + i\psi_{2, i}^* \dpl\psi_{2, i}
        - (\sum_{j=1}^n \psi_{1, i} \psi_{2,j}^* \d_i\d_j W(\vp)
        + {\rm h.c.}) \Big)
        \ ,
        \label{OnShellLagrangian}
\end{eqnarray}

It is clear that the Hamiltonian \eqref{wzham} is of the type
obtained by applying a Legendre transformation to the on-shell
Lagrangian \eqref{OnShellLagrangian}.

\subsection{Topological Twisting}

In this section and the next, we recall the ideas from topological
field theory which are necessary to formulate the $tt^*$ equations
\cite{Cecotti:1991me}. An excellent introductory reference on
topological field theory is \cite{Birmingham:1991ty}.

One may couple a Landau-Ginzburg theory to an arbitrary $U(1)$
gauge field $A_\mu$, so that the correlator $\ex{\prod {\cal O}}$
of an arbitrary product of local operators ${\cal O}$ depends on
$A_\mu$ as well as the spin connection $\omega_\mu$. This coupling
introduces an extra term into the Lagrangian, given by
\begin{equation} \label{eq:extraterms}
    A_z (\psi_-)^i (\psi_-)_i + A_\zbar  (\psi_+)^i (\psi_+)_i.
\end{equation}
With $A$ set equal to $\half$ times the spin connection, a field
which previously had spin $s$ and fermion charge $q$ will now have
spin $s - \half q$. In particular, $Q_+$ which had spin $1/2$ and
fermion number $+1$ becomes a scalar.

With operators ${\cal O}$ set equal to chiral primary fields and
with $A_\mu \sim \f12 \omega_\mu$, the correlator
\[
    \ex{\prod {\cal O}}_{A_\mu, \omega_\nu}
\]
is a topological invariant. In particular one computes from
(\ref{eq:extraterms}) that gauging with the connection
\[
    A_z = -\frac{i}{2} \omega_z, \quad A_\zbar = + \frac{i}{2}
    \omega_\zbar
\]
has the effect of modifying the stress energy tensor,
\begin{equation} \label{SEtensor}
    T_{ab} \ \To \ T_{ab}' \equiv T_{ab} - \f12 {\eps_a}^c \d_c J_b
\end{equation}
The new stress energy tensor is BRST exact, which means that
$T_{ab}' = \{Q, \Lambda_{ab}\}$ for some $\Lambda_{ab}$, where $Q
= Q_+ + \bar{Q}_-$. Observables in the topological theory are
identified with BRST exact objects. Any theory in which the action
is supersymmetric and the stress-energy tensor is a $Q$-commutator
is topological, since by definition the stress-energy tensor
generates metric deformations.

\subsection{The $tt^*$ Equations}

Physical observables of an $N=2$ SCFT are associated with chiral
superfields, with components
\begin{equation}
    \Phi_i =
    ( \phi_{i}^{(0)} (z,{\bar z}) ,
    \phi_{i}^{(1)} (z, {\bar z}),
    {\bar \phi}_{i}^{(1)} (z,{\bar z}),
    \phi_{i}^{(2)}(z, {\bar z}) )
\end{equation}
where
\begin{equation}
    \phi_{i}^{(2)} = \{ Q^- , [ {\bar Q}^- , \phi_{i}^{(0)} ] \}
\end{equation}

We define a deformed theory parameterized by the coupling
constants $(t_i , {\bar t}_{\bar i})$ as follows
\begin{equation} \label{deformed_theory}
    {\cal L} (t_i , {\bar t}_i) = {\cal L}_{0}^{N=2}  + \sum_i t_i
    \int_{\Sigma} \phi_{i}^{(2)}  + \sum_{\bar i} {\bar t}_{\bar i}
    \int_{\Sigma} {\bar \phi}_{\bar i}^{(2)}
\end{equation}
This deformed theory can be transformed into a TFT by the twisting
mechanism. If some of the non-vanishing coupling constants
correspond to relevant deformations, then the theory defined by
\eqref{deformed_theory} will represent a massive deformation of
the $N = 2$ SCFT defined by ${\cal L}_{0}^{N=2}$.

Let $|i,t, {\bar t}; \beta \>$ be the state defined by inserting
on the hemisphere the field $\phi_i$ and projecting to a zero
energy state by gluing the hemisphere to an infinitely long
cylinder of perimeter $\beta$. This corresponds to using a metric
$g = e^{\phi} dz d{\bar z}$ with $\beta = e^{\phi}$. Introduce a
set of connection forms $A_i$, $A_{\bar i}$, defined by
\begin{equation} \label{connec1}
    \< {\bar k} | \d_i - A_i | j \> = 0
\end{equation}
with $|{\bar k} \>$ the antiholomorphic basis. An alternate
definition is in terms of the hemisphere states,
\begin{eqnarray}
    \d_{t_i} |j,t, {\bar t}; \beta \> &=& A_{ij}^{k}
    |k,t,{\bar t}; \beta \>  +  \te{$Q^+$-exact}
    \nonumber \\
    \d_{{\bar t}_i} |j,t, {\bar t}; \beta \>
    &=& A_{{\bar i}j}^{k} \; |k,t,{\bar t}; \beta \>
    + \te{$Q^+$-exact}
    \label{connec2}
\end{eqnarray}
The connection \eqref{connec1} is related to \eqref{connec2} by
\[
    A_{ij}^{k} = A_{ij {\bar k}} g^{{\bar k} k}
\]
with $g^{{\bar k} k}$ the inverse of the hermitian metric
$g_{i{\bar j}} = \< {\bar j} | i \>$.

Therefore the covariant derivatives are given by
\begin{equation}
    D_i
    =
    \d_i - A_i \, , \qquad {\bar D}_{\bar i}
    = \d_{\bar i} - A_{\bar i}
\end{equation}
Using the functional integral representation of $|i,t,{\bar t} \>$
and interpreting the partial derivative $\d_i$ as the insertion
and integration over the hemisphere of the operator
$\phi_{i}^{(2)}$, we conclude by contour deformation techniques
that $\d_i |j,t, {\bar t}; \beta \>$ is also a physical state and
$A_{{\bar i} j}^{k} = 0$.

Defining now
\begin{equation}
    A_{ijk} = \< k | \d_i | j \> = A_{ij}^{l} \; \eta_{lk}
\end{equation}
for $\eta_{lk}$ the topological metric, standard functional
integral arguments give curvature equations for the connections
$A_i$. In particular,
\begin{equation} \label{tts1}
    \d_{\bar l} A_{ij}^{k}
    =
    \beta^2 [ \: C_i , {\bar C}_{\bar l} \: ]_{j}^{k}
\end{equation}
where $C_i$ are chiral ring structure constants
\cite{Lerche:1989uy}. Equation \eqref{tts1}, first written down by
Cecotti and Vafa \cite{Cecotti:1991me}, together with
\be \label{tts2}
    [ D_i, D_j ]
    =
    [ {\bar D}_{\bar i}, {\bar D}_{\bar j} ]
    =
    [ D_i, {\bar C}_{\bar j} ]
    =
    [ {\bar D}_{\bar i}, C_j ]
    = 0
\ee
\be \label{tts3}
    [ D_i, C_i ] = [ D_j, C_i ]
    \; \; ,
    \; \;
    [ {\bar D}_{\bar i}, {\bar C}_{\bar j} ]
    =
    [ {\bar D}_{\bar j} , {\bar C}_{\bar i} ]
\ee
which are deduced by similar techniques, are known as the $tt^*$
equations. To contract the indices of the topological and
antitopological structure constants, one must use the metric $g_{i
{\bar j}}$ of the physical Hilbert space $\cal H$.


\section{Holomorphic Analysis in Banach Spaces}

We now begin the mathematically rigorous portion of the paper. We
briefly review classical theory of analytic mappings between
(possibly infinite dimensional) Banach spaces and some aspects of
Schr\"odinger operators, and prove results which are a
generalization of Kato-Rellich theory. These results are applied
in the next section, which introduces vacuum bundle theory for
Schr\"odinger operators.

\subsection{Holomorphic Families of Unbounded Operators}

A family of bounded operators $T(\chi) \in {\cal B}(X,Y)$ between
two Banach spaces is said to be \emph{holomorphic} if it is
differentiable in norm for all $\chi$ in a complex domain. For
applications, it is not sufficient to consider bounded operators
only, and the notion of holomorphy needs to be extended to
unbounded operators. $D(\,\cdot\,)$ denotes the domain of an
operator, and $\rho(\,\cdot\,)$ denotes the resolvent set.

\begin{definition} \label{def:holo-clsedops}
A family of closed operators $T(\chi) : X \to Y$ defined in a
neighborhood of $\chi = 0$, where $X, Y$ are Banach spaces, is
said to be \emph{holomorphic at $\chi = 0$} if there is a third
Banach space $Z$ and two families of bounded operators $U(\chi) :
Z \to X$ and $V(\chi) : Z \to Y$ which are bounded-holomorphic at
$\chi = 0$ such that $U(\chi)$ is a bijection of $Z$ onto
$D(T(\chi))$, and $T(\chi) U(\chi) = V(\chi)$.
\end{definition}

An equivalent condition which is easier to check in some cases
involves holomorphicity for the resolvent in the usual sense of
bounded operators. We have \cite{K}:

\begin{theorem}  \label{thm:unbdd-ops}
Let $T(\chi)$ be a family of closed operators on a Banach space
$X$ defined in a neighborhood of $\chi = 0$, and let $\zeta \in
\rho(T(0))$. Then $T(\chi)$ is holomorphic at $\chi = 0$ if and
only if for all $\chi$ in some small ball, $\zeta \in
\rho(T(\chi))$ and $R(\zeta, \chi) := (T(\chi) - \zeta)^{-1}$ is
bounded-holomorphic. In this situation, $R(\zeta, \chi)$ is
jointly bounded-holomorphic in two variables.
\end{theorem}

An interesting variant of this (which will arise in the case of
interest for this paper) is the following.

\begin{definition}
A family $T(\chi)$ of closed operators from $X$ to $Y$ defined for
$\chi$ in a domain $\Omega$ in the complex plane is said to be
\emph{holomorphic of type (A)} if
\begin{enumerate}
\item  The domain $D := D(T(\chi))$ is independent of $\chi \in \Omega$.
\item  For every $u \in D$, $T(\chi) u$ is holomorphic for $\chi \in
\Omega$.
\end{enumerate}
\end{definition}

A family that is type (A) is automatically holomorphic in the
sense of Definition \ref{def:holo-clsedops}, taking $Z$ to be the
Banach space $D$ with norm $\norm{u}_Z := \norm{u} +
\norm{T(0)u}$. We now consider analytic perturbations of the
spectrum. The following Theorem of \cite{K} will be used in the
proofs of our main results.

\begin{theorem} \label{lemma:kato-jordan}
Let $X$ be a Banach space and $T(\chi) \in \calC(X)$ be
holomorphic in $\chi$ near $\chi = 0$ and let $\Sigma(0) =
\Sigma(T(0))$  be separated into two parts $\Sigma'(0),
\Sigma''(0)$  in such a way that there exists a rectifiable simple
closed curve $\Gamma$ enclosing an open set containing $\Sigma'$
in its interior and $\Sigma''$ in its exterior.  In this
situation, for $\abs{\chi}$ sufficiently small, $\Sigma(T(\chi))$
is also separated by $\Gamma$ into two parts $\Sigma'(T(\chi))
\cup \Sigma''(T(\chi))$, and $X$ decomposes as a direct  sum $X =
M'(\chi) \oplus M''(\chi)$ of spectral subspaces. Moreover, the
projection on $M'(\chi)$ along $M''(\chi)$ is given by $P(\chi) =
-\frac{1}{2\pi i} \oint_\Gamma R(\zeta, \chi) d\zeta$ and is
bounded-holomorphic near $\chi = 0$.
\end{theorem}

\begin{remark}
The projection $P(\chi)$ is called the {\it Riesz projection},
and this projection being bounded-holomorphic is equivalent to the
statement that the subspaces $M'(\chi)$ and $M''(\chi)$ are
holomorphic in their dependence on $\chi$.
\end{remark}

\subsection{Perturbation Theory}

Consider a family of closed operators $T_\chi$ depending on a
parameter $\chi \in B_\eps(0)$ for some $\eps > 0$, with a common
domain $D$ in a Hilbert space $\H$, and such that each $T_\chi$
has a nonempty resolvent set. Write $T_\chi = T_0 +
V_{\te{eff}}(\chi)$, where $V_{\te{eff}}(\chi) := T_\chi - T_0$ is
called the effective potential.

\begin{definition} \label{def:eigenstability}
A discrete eigenvalue $\lambda$ of $T_0$ is said to be
\emph{stable} with respect to $V_{\te{eff}}$ if
\begin{enumerate}
\item  $\exists \, r > 0$ s.t. $\Gamma_r \equiv \{ |z-\lambda| =
r\} \subset \rho(T_\chi)$ for all $|\chi|$ sufficiently small, and

\item $P(\chi) \equiv -\frac{1}{2\pi i} \oint_{\Gamma_r}
(T_\chi - \zeta)^{-1} d\zeta$ converges to $P(0)$ in norm as $\chi
\to 0$.
\end{enumerate}
\end{definition}

The notion of stability arises in the following rigorous statement
of degenerate perturbation theory, due in its original form to
Kato. Here, $m(\lambda)$ denotes multiplicity of eigenvalue
$\lambda$.

\begin{theorem}[Degenerate Perturbation Theory]
\label{thm:degen-ptheory} Let $T_\chi$ be a Type (A) family near
$\chi_0 = 0$, and let $\lambda_0$ be a stable eigenvalue of $T_0$.
There exist families $\lambda_\ell(\chi), \ \ell = 1 \ldots r,$ of
discrete eigenvalues of $T_\chi$ such that
\begin{enumerate}
\item  $\lambda_\ell(0) = \lambda_0$ and $\sum_{\ell = 1}^r
m(\lambda_\ell(\chi)) = m(\lambda_0)$.
\item Each $\lambda_\ell(\chi)$ is analytic in $\chi^{1/p}$ for
some $p \in \Z$, and if $T_\chi$ is self-adjoint $\forall \chi \in
\R$, then $\lambda_\ell(\chi)$ is analytic in $\chi$.
\end{enumerate}
\end{theorem}


\section{The Vacuum Bundle for Schr\"odinger Operators} \label{sec:schrodinger}

The free Schr\"odinger operator $P^2 = -\Delta$ in $d$ space
dimensions is self-adjoint on the domain $D(P^2) = H^2(\R^d)$, and
has $C_0^\infty(\R^d)$ as a core. We consider perturbations
$V_\chi$ of $H_0$ which depend analytically on (coupling)
parameters $\chi$, and show that for certain reasonable classes of
such perturbations, the total Schr\"odinger operator $P^2 +
V_\chi$ remains self-adjoint and has the appropriate spectral
splitting condition to apply holomorphic Kato theory. Ultimately
this leads to the rigorous construction of a vacuum bundle for
quantum mechanics, which is used later for vacuum estimates in the
more complicated Wess-Zumino field theory model.

There are a number of conditions on a potential $V$ which
guarantee that the Schr\"odinger operator $P^2 + V$ will be
essentially self-adjoint. An example on $\R^3$ of one such
condition is the following. Let $R$ denote the family of
potentials $f(x)$ on $\R^3$ obeying
\[
    \int \f{\abs{f(x)} \, \abs{f(y)}}{\abs{x-y}^2} \, dx dy
    < \infty.
\]
Then $V \in L^\infty(\R^3) + R \ \Rightarrow\ P^2 + V$ is
essentially selfadjoint \cite{Simon}.

\begin{definition}
We will refer to a function space $W$ as a \emph{space of
admissible potentials} if $\forall\ f \in W$, the Schr\"odinger
operator $P^2 + f$ is essentially self-adjoint.
\end{definition}

Standard self-adjointness theorems for Schr\"odinger operators
have the property that the space $W$ of all admissible potentials
is a locally convex space. A locally convex topological vector
space is the minimal structure which is necessary for the
traditional definition of ``holomorphic map'' to remain valid with
no modifications. A map $T : U \to W$ from a domain $U \subset \C$
into a locally convex space $W$ is said to be {\it holomorphic} at
$z_0 \in U$ if $\lim_{z \to z_0} \frac{T(z) - T(z_0)}{z - z_0}$
exists. These definitions allow us to speak of a holomorphic map
$V$ from a complex manifold $M$ into a space $W$ of admissible
potentials. This generalizes the notion of a perturbation which
depends on a number of coupling parameters; in our case
coordinates on a manifold $M$ take the role of generalized
couplings.

\begin{theorem} \label{thm:perturbation}
Let $U \subset X$ be an open connected set in a Banach space $X$
and let $\H$ be a Hilbert space. Let $H_0$ be a closed operator
on a dense domain $\D \subseteq \H$. Fix a map $V : U \to
Op(\H)$, and for $\tau \in U$, define $H(\tau) = H_0 + V(\tau)$,
which we assume has nonempty resolvent set. Assume $\forall \tau,
\, V(\tau)$ has $H_0$-bound smaller than one, and that $V(\tau)
\psi$ is analytic in $\tau$, for any $\psi \in \D$. Then $H(\,
\cdot \,)$ is analytic.
\end{theorem}

\prooftheorem{thm:perturbation} By the Kato stability theorem
\cite{K}, $H(\tau)$ is closed for all $\tau$. Since $V(\tau)$ is
$H_0$-bounded, $D(H(\tau)) = D(H_0) \cap D(V(\tau)) = D(H_0)$. It
follows that the family $H(\,\cdot\,)$ is type (A), and hence
analytic. \endproof

\begin{remark}
Our assumptions in Theorem \ref{thm:perturbation} are sufficiently
general to allow the domain of the map $V$ to be an arbitrary
manifold.
\end{remark}

If we assume $H_0$ to be a selfadjoint operator on a dense domain
$\D \subset L^2(\R^n)$, and we let $V_i$ for $i \in \N$ be a
sequence of uniformly bounded operators on $L^2(\R^n)$ and $\tau
\in \ell^\infty(\C)$, then Theorem \ref{thm:perturbation} implies
that the Hamiltonian
\[
    H(\tau) = H_0 + \sum_{i=1}^\infty  \tau_i V_i
\]
is analytic in the coupling parameters $\tau_i$.

In order to apply the Remark following Theorem
\ref{lemma:kato-jordan}, we need to work in a scenario where the
lowest eigenvalue of the Schr\"odinger operator is an isolated
eigenvalue. This is by no means guaranteed; in fact it is
typically false on $L^2(\R^n)$ when $V(x)$ is continuous and
$\lim_{x\to \infty} V(x) = 0$. However, this spectral gap is
guaranteed given a compact manifold and some very generic
conditions on $V$, and on a noncompact manifold such as $\R^d$
when $V(x)$ grows at infinity. We discuss both the compact and
non-compact cases since the non-compact case is usually studied in
quantum mechanics, but quantum field theory is frequently studied
on a compact manifold.

\begin{lemma} \label{lemma:cpct_manifold}
Define $H = -\Delta + V(x)$ on $L^2(X)$ for a compact Riemannian
manifold $X$, and assume that $V \in L^2(X)$ with $V(x) \geq 0$.
Then $H$ has purely discrete spectrum in which the eigenvalues are
not bounded above, and all eigenvalues have finite algebraic
multiplicity.
\end{lemma}

The proof of Lemma \ref{lemma:cpct_manifold} uses standard methods
along the lines of Griffiths and Harris' proof of the Hodge
theorem \cite{GH}. Lemma \ref{lemma:cpct_manifold} implies a
spectral gap between the lowest eigenvalue (ground state) of $H$
and the first excited state eigenvalue on a compact manifold.

Generally, if the resolvent $R_H(z)$ is compact, then
$\sigma(R_H(z))$ is discrete with 0 the only possible point in
$\sigma_{ess}$. Hence one would expect that $H$ has discrete
spectrum with the only possible accumulation point at $\infty$,
and this implies $\sigma_{ess}(H) = \emptyset$. This reasoning
shows that if $V(x) \geq 0$, $V$ is in $C(\R^n)$ or
$L^2_{loc}(\R^n)$, and $V(x) \to \infty$ as $\norm{x} \to \infty$,
then $H = -\Delta + V$ has purely discrete spectrum on
$L^2(\R^d)$.

If $z_0 \in \sigma(T_0)$ is an $N$-fold degenerate eigenvalue of
$T_0$, then generically a perturbation will break the degeneracy,
and therefore, if $T_\kappa$ is a holomorphic perturbation of
$T_0$ we expect, as in Theorem \ref{thm:degen-ptheory}, a number
of eigenvalue curves which flow away from $z_0$. It follows that
we have a vacuum bundle only in the special cases when the
$N$-fold degeneracy is not broken by $T_\kappa$, for all $\kappa$
lying in some complex manifold. Physics intuition suggests the
only way this can happen is in the presence of additional
symmetry, such as supersymmetry. In the latter case, the Witten
index \cite{Witten:df}, which counts the ground states weighted by
$(-1)^F$, equals the index of the Dirac operator and this does not
change for all effective superpotentials of the same degree.

The following is the main theorem of Section
\ref{sec:schrodinger}. It asserts the existence of the vacuum
bundle for a Schr\"odinger operator.

\begin{theorem}  \label{thm:schrod}
Let $M$ be either a finite dimensional complex manifold or an
infinite-dimensional complex Banach manifold, and let $X$ be a
finite dimensional real manifold with a Riemannian metric. Let $Y$
be a linear space whose elements are complex-valued functions on
$X$, such that for any $f \in Y$,
\begin{enumerate}
\item  The multiplication operator $f$ on $L^2(X)$
is $P^2$-bounded with $P^2$-bound $< 1$.
\item $P^2 + f$ has spectral gap between first and second
eigenvalues on $L^2(X)$.
\end{enumerate}
Let $V : M \to Y$ be holomorphic, and for $\tau \in M$ let
$H_\tau := P^2 + V_\tau$ have lowest energy $\lambda_0(\tau)$
with eigenspace $E_0(\tau)$. If $\dim E_0(\tau)$ is constant,
then $E_0 \to M$ is a holomorphic vector bundle.
\end{theorem}

\prooftheorem{thm:schrod} Since $V(\tau)$ is $P^2$-bounded with
$P^2$-bound $< 1$, the Kato-Rellich theorem implies that for any
$\tau \in M$, $H(\tau) = P^2 + V(\tau)$ is self-adjoint on $\D =
D(P^2)$. For $\psi \in \D$, $V(\tau) \psi$ has the form of
$f(\tau,x)\psi(x)$ as a function on $x \in X$, where $f(\tau,x)$
is analytic in $\tau$ by assumption. We conclude by Theorem
\ref{thm:perturbation} that $H(\tau)$ is analytic. To show that
the ground state subspace is analytic, we work with operators
having discrete spectrum with spectral gap (see Lemma
\ref{lemma:cpct_manifold} and the discussion thereafter). We may
therefore apply the Remark following Lemma
\ref{lemma:kato-jordan}. Since $\dim E_0(\tau)$ is constant, it
follows that we may choose $N$ holomorphic functions $v_i(\tau),
i=1\ldots N$ s.t. $\forall \, \tau$,  $\{v_i(\tau)\}$ form a
linearly independent spanning set of $E_0(\tau)$.
\endproof

\begin{remark} The space $Y$ has to be tuned to the space $X$ so that
conditions 1 and 2 in the theorem are satisfied. For example, if
$X = \R^d$, then $Y$ can consist of elements of $C(\R^d)$ or
$L^2_{loc}(\R^d)$ that blow up at infinity. If $X$ is a compact
manifold, then we can take $Y = \{ f \in L^2(X) : f(x) \geq 0 \
\forall x \in X\}$. This suggests a general class of new problems
in functional analysis. Given $X$, the problem is to determine the
largest space $Y$ which is tuned to $X$ in the sense of Theorem
\ref{thm:schrod}.
\end{remark}

\section{The Wess-Zumino Model, the Dirac Operator on Loop Space,
and Vanishing Theorems}  \label{sec:wzmodel}

\subsection{The Wess-Zumino Model on a Cylinder}

In a fundamental paper \cite{JLW}, Jaffe, Lesniewski, and Weitsman
present rigorous results for supersymmetric Wess-Zumino models by
generalizing index theory of Dirac operators to an infinite
dimensional setting; we now give a concise introduction to the
results of \cite{JLW} and recall a number of facts from
constructive field theory which will be needed in later sections.

We study self-adjoint Hamiltonians $H$ defined on the Hilbert
space $\H = \H_b \otimes \H_f$, where $\H_b$ and $\H_f$ are,
respectively, the symmetric and antisymmetric tensor algebras
over the one-particle space $W = W_+ \oplus W_-$, where $W_+$ and
$W_-$ represent single particle/antiparticle states respectively,
and $W_\pm \equiv L^2(T^1)$. The Hamiltonian is that
corresponding to one massive complex (Dirac) fermion field $\psi$
of mass $m$, and one complex boson field $\vp$ with the same mass
as the fermion field, defined on a circle of length $\ell$. The
interactions are parameterized by a holomorphic polynomial
$V(z)$, known as the {\it superpotential}. The free Hamiltonian
in second-quantized notation is written as
\[
H_0 = \sum_{j=\pm, \ p \in \hat T^1} \omega(p) \left( a_j^*(p)
a_j(p) + b_j^*(p) b_j(p) \right) \, ,
\]
where $a_j$ satisfy canonical commutation relations for bosonic
oscillators, and $b_j$ satisfy the corresponding Fermion algebra.

We can write the superpotential as $V(\vp) = \half m \vp^2 +
P(\vp),$ separating out the mass term.  The energy density of the
bosonic self interaction is $\abs{\d V(\vp)}^2$, a polynomial of
degree $2n-2$. The boson-fermion interaction is known as a
generalized Yukawa interaction, and has the form
\[
    \bar\psi \Lambda_+ \psi \d^2 V + \bar\psi \Lambda_- \psi (\d^2 V)^*,
\]
where $\Lambda_\pm$ are projections onto chiral subspaces of
spinors. If $P=0$, this interaction reduces to a free mass term
$m\bar\psi \psi$.

Define operators $N_{\tau, \{b,f\}}$ by
\[
N_{\tau, b} = \sum_{j=\pm, p \in \hat T^1} \omega(p)^\tau
a_j^*(p) a_j(p), \qquad N_{\tau, f} = \sum_{j=\pm, p \in \hat
T^1} \omega(p)^\tau b_j^*(p) b_j(p).
\]
Then the family of operators $N_\tau = N_{\tau,b} \otimes I + I
\otimes N_{\tau,f}$ interpolates between the total particle number
operator $N_0$ and the free Hamiltonian $N_1$. We write $N_f$ for
$N_{0,f}$. A selfadjoint unitary operator that is not the identity
necessarily has $+1$ and $-1$ eigenvalues, and is therefore a
$\Z_2$-grading. $\Gamma = \exp(i \pi N_f)$ is self-adjoint and
unitary, hence the Hilbert space splits into a direct sum $\H =
\H_+ \oplus \H_-$ of the $\pm 1$ eigenspaces of $\Gamma$, and thus
naturally inherits the structure of a super vector space.

The following bilinear form over $\H$ is known as the {\it
supercharge}:
\begin{equation} \label{eq:dirac-op}
Q = \frac{1}{\sqrt{2}} \int_{T^1} dx \ \psi_1(\pi - \d_1 \vp^* -
i \d V(\vp)) + \psi_2 (\pi^* - \d_1 \vp - i \d V(\vp)^* ) + h.c.
\end{equation}
where the superpotential $V(\vp) = \f12 m \vp^2 + \sum_{j=3}^n a_j
\vp^j$ is a holomorphic polynomial with $n \geq 3,\ a_n \ne 0$,
and $m > 0$. With appropriate regularization and limiting
procedures, we have $H = Q^2$, where $H$ is the full interacting
Hamiltonian.

Define $\D(T^1)$ as the space of smooth maps $T^1 \to \C$, with
topology defined by uniform convergence of each derivative.
$\D(T^1)$ is an infinite-dimensional Fr\'echet manifold known as
{\it loop space}, and $Q$ has the structure of a Dirac operator on
loop space. The proof that the bilinear form \eqref{eq:dirac-op}
defines an operator requires careful analysis, which has been done
in \cite{JLW}. The strategy is to split the expression
\eqref{eq:dirac-op} for $Q$ into a free part and an interacting
part, and to further regularize the interacting part by convolving
the fields $\vp(x), \psi_\mu(x)$ with a smooth approximation to
the periodic Dirac measure, which implements a momentum space
cutoff.

To obtain the desired approximation to periodic Dirac measure, we
use a cutoff function $\chi$ satisfying: $\ $  $0 \leq \chi \in
{\mathcal S}(\R)$, $\ $ $\int_{-\infty}^\infty \;\chi(x) dx =1$,
$\ $ $\chi(-x)=\chi(x)$, $\ $  $\hat \chi(p) \geq 0$, $\ $ ${\rm
supp} \;\hat\chi(p) \subset [-1,1]$, $\ $ and  $\ \hat\chi(p)
> 0$ \ for \ $|p| \leq 1/2$. We set
\[
    \chi_\kappa(x) = \kappa \sum _{n \in \Z}
    \chi(\kappa(x-n\ell))
\]
where $\kappa > 0$. Regularized (cutoff) fields are defined by
taking convolution with $\chi_\kappa$ on $T^1$,
\[
    \vp_\kappa(x) =  \chi_\kappa \ast \vp (x),
    \ \
    \psi_{\mu ,\kappa}(x) = \chi_\kappa \ast \psi_\mu (x)\, .
\]
The result of this procedure is a regularized supercharge
$Q(\kappa) = Q_0 + Q_{i,\kappa}$. {\it A priori} estimates
\cite{2DWZW} establish a homotopy between $Q(\infty)$ and $Q(0)$
with $i(Q_+(\kappa))$ constant. Explicit calculation\cite{JLW}
shows that $Q_0^0 + Q_{i,0}$ is the supercharge of the model of
$N=2$ holomorphic quantum mechanics considered in \cite{JLL} and
this paper. Existence of a holomorphic vacuum bundle for the
quantum mechanical supercharge $Q_0^0 + Q_{i,0}$ follows by
dimensional reduction from Theorem \ref{thm:grstates-wzmodel}.
However the holomorphic quantum mechanics model is sufficiently
simple that the desired vacuum bundle estimates can be established
directly using methods of classical ODEs, as we show in Section
\ref{sec:qm}.

It was shown in \cite{JLL} that $Q(0)$ has only bosonic ground
states, i.e. $n_-(Q(0)) = 0$. We say that a Hamiltonian has the
{\it vanishing property} if $n_- = 0$.

\subsection{The $N=2$ Wess-Zumino$_2$ Vanishing Theorem}

We recall the vanishing theorem for the $N=2$ Wess-Zumino model
defined on a cylindrical spacetime of perimeter $\ell$,
independently proven by Janowsky and Weitsman \cite{JW}, and by
Borgs and Imbrie \cite{Borgs:gq}, which is crucial for later
sections. Consider superpotentials of the form
\be \label{eq:JWsuper}
    V = \lambda^{-2} \t{W}(\lambda x) + \lambda^{-1} \xi w(\lambda x)
\ee
where $\t{W}$ and $w$ are polynomials of degree $n$, $\t{W}'$ has
$n-1$ distinct zeros, and $|\t{W}''| = 1$ at each zero. For $\ell
> 1$, the $N=2$ Wess-Zumino quantum field theories corresponding to
superpotentials of type \eqref{eq:JWsuper} have no fermionic zero
modes for $\lambda$ and $\xi$ sufficiently small, where $\lambda$
is a parameter that controls the depth and spacing of the
potential wells, and $\xi$ measures the strength of $w$, which
represents a small perturbation away from the unit mass condition.

To see this, we note that results of \cite{2DWZW,JL} imply that
$e^{-\tau H}$ is trace class for all $\tau > 0$ and $\ind(Q) =
\tr(\Gamma e^{-\tau H}) = \deg(V)-1$. It follows that
\[
    \dim \ker H = \lim_{\tau \to \infty} \Tr(e^{-\tau H}),
\]
and given the assumptions on $\lambda$ and $\xi$, cluster
expansion methods (Theorem 3 of \cite{JW}) show that for all
$\ell$ larger than some constant, there exists  $\tau_\ell$
sufficiently large so that
\be \label{eqn:diff-tr}
    \abs{\Tr(\Gamma e^{-\tau_\ell H})
        - \Tr(e^{-\tau_\ell H})
        }
    <
    \f12 \, .
\ee
The condition on $\ell$ is necessary because the proof of
Janowsky-Weitsman Theorem 3 proceeds by an estimate of the form
\be \label{thm3}
    \exp[ -c_1 \tau + c_2 \ell \tau e^{-\ell}] < \f12
\ee
where $c_1, c_2$ are constants. As long as $\ell e^{-\ell} <
c_1/c_2$, we can always find $\tau$ such that \er{thm3} holds, but
if $\ell e^{-\ell} \geq c_1/c_2$, there is no acceptable $\tau$.

Now $Q$ is selfadjoint, $H = Q^2 \geq 0$ and \cite{JL} shows that
$e^{- \tau H}$ is trace class, hence
\[
    \dim\ker H = \Tr(e^{-\tau H}) + O(e^{-\tau \eps})
\]
for $\tau \gg 1$ and for some $\eps > 0$. It now follows from
\eqref{eqn:diff-tr} that
\[
    \abs{\dim\ker H - \ind(Q)} < 1.
\]
In this situation, $\dim\ker H$ and $\ind(Q)$ are integers
differing by less than one, hence they are equal. It follows that
for superpotentials as in \eqref{eq:JWsuper}, $n_-(H) = 0$.

The vanishing theorem stated above for weakly coupled $N=2$
Wess-Zumino models also follows directly from Theorem 2.2 of Borgs
and Imbrie \cite{Borgs:gq}, which assumes that the cylinder size
$L$ is greater than 1. In either case, a condition of the form $L
> const$ is required.

\subsection{Other vanishing theorems}
Some care is required, as the term `vanishing theorem' can take on
other, perhaps contradictory, interpretations. For example, if $M$
is a compact spin manifold with a nontrivial $S^1$-action, Atiyah
and Hirzebruch \cite{AH} have shown that $\Ind(D) = \hat A(M) =
0$, where $D$ is the Dirac operator on $M$. In a situation more
closely related to quantum field theory, Witten
\cite{Witten:1987cg} formally applied the Atiyah-Bott-Segal-Singer
fixed point formula to the Dirac operator $D^L$ on loop space
$LM$, with the result that, with $M$ as above and under suitable
assumptions on the first Pontryagin class, the Witten genus
$\Ind(D^L)=0$. In the present context, $\Ind Q = 0$ would give the
false conclusion $n_-(Q) = n_+(Q)$, and does not imply that the
zero modes are purely bosonic, so the Janowsky-Weitsman and
Borgs-Imbrie theorem is a qualitatively different result from
Witten's vanishing theorem. In fact $\Ind Q \ne 0$ for the $N=2$
Wess-Zumino models, so Witten's result does not apply at all.

We will show that the vacuum bundle exists for $N=2$ models with
the vanishing property. A large class of Wess-Zumino models
(precisely those with superpotentials of the form
\eqref{eq:JWsuper}), are known to have the $n_- = 0$ property. We
conjecture that a vanishing theorem stronger than \cite{JW} holds,
and that all $N = 2$ Wess-Zumino models on a cylinder satisfy $n_-
= 0$.

It is interesting to note that the vanishing theorem of
Janowsky-Weitsman \cite{JW} and Borgs-Imbrie \cite{Borgs:gq} is
expected \emph{not} to hold for the corresponding $N=1$
Wess-Zumino models. Jaffe {\it et al} \cite{JLL} considered a
quantum mechanics version of the \ $N=1$ \ Wess-Zumino field
theory. Supersymmetry is broken or unbroken depending on the
asymptotics of the superpotential at infinity, and is
characterized by its degree:
\[
    i(Q_+)=\pm \;{\rm deg} \; V \ \ (\te{mod}\ 2).
\]
In the unbroken case, there is a unique ground state; it belongs
to \ ${\cal H}_+ \ (n_+=1, \ n_-=0)$ \ or to \ ${\cal H}_- \
(n_+=0, \ n_-=1),$ according to the additional $\Z_2$ symmetry of
the superpotential. In the case of broken supersymmetry, there are
exactly two ground states and \ $n_+ = n_- = 1$.  Similar results
are true in the corresponding \ $d=2$ \ quantum field models in a
finite volume \cite{JLW}.

Thus the vanishing property is an aspect of $N=2$ supersymmetry,
as is the theory of the ground state metric, $tt^*$ geometry, and
the CFIV index \cite{Cecotti:1992qh,Cecotti:1991me}.

\subsection{The Vacuum Bundle and Atiyah-Singer Index Theory}
\label{sec:indextheory}

Let $\calC(\H)$ denote the space of closed unbounded operators on
Fock space $\H = \H_b \otimes \H_f$. Suppose that we have
identified the appropriate moduli space $\M$ of coupling constants
for a supersymmetric quantum field theory with supercharge $Q$ and
Hamiltonian $H$. For example, the space $\JW$ introduced our
construction of the vacuum bundle is such a space (although not
the largest) for $N=2,\ n_- = 0$ Wess-Zumino theories.

In view of the theory developed in Sections \ref{sec:wzmodel} and
\ref{sec:modulispace}, quantum field theory provides a map from
the total moduli space $\M$ into $\calC(\H)$, given by associating
the supercharge operator $Q_{\T}$ to any set of coupling constants
$\T \in \M$. Composing this map with the squaring function gives
the Hamiltonian of the theory also as a map $\M \To \calC(\H)$,
defined by $\T \to (Q_{\T})^2 \equiv H_{\T}$. This induces a map
from $\M \to Gr(\H)$ given by associating $\T \to \ker H_{\T}$,
where $Gr(\H)$ denotes the Grassmannian of closed subspaces of
$\H$, with topology given by identifying closed subspaces with
projectors and imposing a standard operator topology.

The vanishing property is the statement that
\[
    \dim \ker \at{H_{\T}}{\H_-} = 0
    \ \text{ for all } \
    \T \in \M
\]
where $\H_-$ denotes the $-1$ eigenspace (or \emph{fermionic
subspace}) of the $\Z_2$-grading operator $\Gamma$.

Let $D : \Gamma(E) \To \Gamma(F)$ be an elliptic operator and let
$E$ and $F$ be vector bundles over a closed manifold $M$. The
Atiyah-Singer Index Theorem states that
\[
    \Ind D := \dim\Ker D - \dim \Coker D
    = \left\< P(M, \sigma_{\rm top}(D)), [M] \right\>.
\]
The quantity on the right is a characteristic number built from
the topology of $M$ and topological information contained in the
top order symbol of $D$.

Atiyah and Singer also proved the Families Index Theorem, which
applies to a family of elliptic operators $D_n$ for $n$ ranging in
a compact manifold $N$. The Families Index Theorem identifies the
Chern character of the index bundle $\uInd(D)$ in $H^*(N; \Q)$
with a characteristic class on $N$ built from the topology of $N$
and the pushforward of the symbols of the operators $D_n$. The
index bundle is a virtual bundle whose fiber for generic $n \in N$
is the formal difference $\Ker(D_n) - \Coker(D_n)$, i.e.
\[
    \uInd D = \uKer(D) - \uCoker(D)
\]
In our framework, $N$ is identified with $\M$, the moduli space of
theories, and each theory $n \in N$ has a supercharge $D_n$.
$\Coker(D_n)$ is then identified with the fermionic zero modes.
Therefore, in supersymmetric quantum field theories with the
vanishing property, $\Coker(D_n) = 0$ for all $n \in \M$ and
index bundle is $\uInd(D) = \uKer(D)$ which is the vacuum bundle.

The Families Index Theorem suggests that the vacuum bundle exists
for supersymmetric theories whenever the following conditions are
satisfied: (a) a compact manifold $\M$ can be identified with
(possibly a subset of) the Moduli space, (b) the vanishing
property holds at every point $\T \in \M$, and (c) the supercharge
$Q_{\T}$ is a closed, densely defined Dirac-type elliptic operator
with $(Q_{\T})^2 = H_{\T}$. We give an existence proof in the next
section that does not rely directly on the index theorem.

\section{Construction of the Vacuum Bundle}
\label{sec:construction}

In this section we give the ground states of the Wess-Zumino
models considered above a geometrical structure, by first
constructing the moduli space of admissible superpotentials (the
\emph{base space} of the vector bundle), and then proving that the
ground states vary holomorphically over this space.

\subsection{The Base Space}
\label{sec:modulispace}

In this section we give a detailed description of the
Janowsky-Weitsman moduli space, showing it to be a differentiable
manifold, and therefore of suitable character to function as the
base space for a vector bundle.

The polynomial superpotential is $W_\lambda(x) \equiv \lambda^{-2}
\t{W}(\lambda x)$, with the assumptions
\begin{enumerate}
\item[(A)] $\t{W}'$ must have $n-1$ distinct zeros, where $n =
\deg\t{W}$, and

\item[(B)] $\abs{\t{W}''} = 1$ at each zero of
$\t{W}'$.
\end{enumerate}
The first condition is motivated by the fact that the bosonic
potential $\abs{W_\lambda'(\phi)}^2$ has minima where $\t{W}'$ has
zeroes. Scaling $\lambda \to 0$ increases the distance between and
the depth of the potential wells. Roughly speaking, the moduli
space of theories we will consider is the space of potentials
satisfying (A) and (B). Such potentials exist; a one-parameter
family with degree $2n'$ is given for $ \beta \in (0,1)$ by
\[
    \t{W}_\beta'(z)
    =
    \prod_{k=1}^{n'-1} \left( 2 \sin \f{\pi k}{n'}
    \right)^{-1} \prod_{k=1}^{n'}
    \left[
    \frac{(z-e^{2\pi i k/n'})(z-e^{2\pi i(k+\beta)/n'})}{2\sin\f{\pi(k+\beta)}{n'}}
    \right]
    \,.
\]
The existence of such families suggests that the space of
superpotentials is a topological space containing continuous
paths. In fact, the space of potentials satisfying (A) has a very
natural geometry; and the restriction (B) will be removed by a
mass perturbation which we will analyze.

We let $\C[X]_n$ denote the set of all polynomials of degree $n$
in one variable over $\C$. We let ${\cal Z}_{n,k}$ denote the
space of all $p(X) \in \C[X]_n$ s.t. $p$ has exactly $k$ distinct
zeros. Also let $P(n,k)$ denote the number of partitions of $n$
with length $k$ and no zero entries. For $1 < k < n$, the space
${\cal Z}_{n,k}$ has $P(n,k)$ distinct connected components, but
for $k = n$ (the case of our interest), the polynomial is uniquely
determined by the $n$ distinct zeros, together with an overall
scaling factor. Therefore,
\be  \label{Znn}
    {\cal Z}_{n,n} =
    \C \times
    \Big\{ (z_1, \ldots, z_n) : z_i \ne z_j \ \forall i,j \Big\}
\ee
In particular, \eqref{Znn} shows that ${\cal Z}_{n,n}$ is
$\C^{n+1}$ minus a closed set, and therefore a differentiable
manifold.

In the case of the Janowsky-Weitsman space, we
need to characterize the set of possible $\t{W} \in \C[z]_n$ such
that $\t{W}' \in {\cal Z}_{n-1,n-1}$. Quite generally, if $S
\subset \C[z]$ is a finite-dimensional manifold, we define the
notation
\be \label{integral}
    \int S
    \equiv
    \{ f(z) \in \C[z] : f'(z) \in S\}\, .
\ee
Then there is a bijective mapping $\int S \longleftrightarrow \C
\times S$ given by mapping the pair $(c, g(z)) \in \C \times S$ to
the polynomial $c + \int_0^z g(w)dw$. The space $\int S$ inherits
the structure of a differentiable manifold in the natural way by
declaring that this bijection is a diffeomorphism.

We conclude that condition (A) is equivalent to the statement:
\[
    \t{W} \in \int {\cal Z}_{n-1,n-1} \, .
\]

The second condition (B) is more problematic because it states
that $(\forall \, i) \, \t{W}''(z_i) \in S^1$, and $S^1$ is a
closed set in $\C$. This problem is resolved by noting that the
results of Janowsky-Weitsman are invariant under perturbations of
the form
\begin{equation} \label{eq:massperturbation}
W_\lambda(x) = \lambda^{-2} \t{W}(\lambda x) + \lambda^{-1}
\epsilon w(\lambda x)
\end{equation}
where $w$ is also a polynomial of degree $n$ and $\epsilon$ is a
small parameter. This breaks any artificial symmetry due to the
mass restriction (B). We wish to analyze this symmetry breaking
and the effect on the masses in greater detail. In order to do
this, we establish that adding a small perturbation to a
polynomial with its zeros separated causes each mass
$\t{W}''(z_k)$ to be perturbed within a similarly small
neighborhood of its unperturbed value. We call this \emph{fine
tuning} of the zeros.


Consider the problem of defining a function $w = f(z)$ by solution
of the algebraic equation $G(w,z) = 0$ where $G$ is an irreducible
polynomial in $w$ and $z$. If $G$ is arranged in ascending powers
of $w$, this equation can be written
\begin{equation} \label{eq:powers}
g_0(z) + g_1(z) w + \dots + g_m(z) w^m = 0
\end{equation}
If we imagine a particular value $z_0$ to be substituted for $z$,
we have an equation in $w$ which, in general, will have $m$
distinct roots $w_0^{(1)}, w_0^{(2)}, \ldots, w_0^{(m)}$. An
exception takes place if and only if
\begin{enumerate}
\item[{\it (i)}] $g_m(z_0) = 0$, in which case the degree of the equation
is lowered, or

\item[{\it (ii)}] $G(z_0, w) = 0$ has multiple roots.
\end{enumerate}

The second case can occur if and only if the discriminant, which
is an entire rational function of the coefficients, vanishes. If
$G(z,w)$ is irreducible, then the discriminant $D(z)$ does not
vanish identically but is a polynomial of finite degree. Thus the
exceptions {\it (i)} and {\it (ii)} can occur for only a finite
number of special values of $z$, which we denote by $a_1, a_2,
\ldots, a_r$, and which we call \emph{excluded points}.

By the implicit function theorem, for any non-excluded $z_0$,
there are $n$ distinct function elements $\omega_1, \ldots,
\omega_n$ such that
\begin{equation} \label{fnelement}
G(z, \omega_j(z)) = 0 \,.
\end{equation}
If we continue one of these function elements $\omega_j$ to
another non-excluded point $z_1$, we get another function element
(over $z_1$) that satisfies \eqref{fnelement}. In this way, the
equation $G(z,w) = 0$ defines a multi-valued function, or Riemann
surface; we state this as a lemma.

\begin{lemma} \label{Riemann}
In the punctured plane
\[
H = \C \setminus \{ a_1, \ldots, a_r \}
\]
the equation $G(z,w) = 0$ defines precisely one $m$-valued regular
function $w = F(z)$.
\end{lemma}

Lemma \ref{Riemann} and the discussion preceding it apply to the
special case in which all but one of the functions $g_i(z)$,
defined in eq.~\eqref{eq:powers}, are constant,
\[
g_i(z) = \begin{cases} c_i, & i \ne k \\ z, &  i = k
\end{cases} \, , \quad c_i \in \C
\]
Away from the excluded points $\{ a_\nu \}$ associated to this
choice, the zeros of $\sum_{i=0}^m g_i(z) w^i$ are distinct and
vary as analytic functions of the coefficient of $w^k$. Repeating
this procedure for each $k=1 \ldots m$, we conclude that away from
excluded points, the zeros depend holomorphically on each
coefficient.

We now reformulate this result in a way that is relevant to
quantum field theory, which we state as Theorem \ref{thm:fine}.
For a polynomial $w(x) = \sum a_i x^i$, we define $\norm{w(x)}^2 =
\sum |a_i|^2$, which gives $\C[x]_n$ the topology of Euclidean
space.

\begin{theorem}[Fine Tuning] \label{thm:fine} Consider a
fixed polynomial superpotential $\t{W}(x)$. Let ${\cal N}$ be a
neighborhood of $0$ in the space $\C[x]_n \cup \{0\}$. Let $Z =
\{\xi_1, \ldots, \xi_n\}$ be the zero set of $\t{W}'(x)$, which we
assume is nondegenerate, and let $Z_{w}$ denote the zero set of
$\f{d}{dx}(\t{W}(x) + w(x))$. For ${\cal N}$ sufficiently small,
we assert that the union $\bigcup_{w \in {\cal N}} Z_w$ takes the
form $\bigcup_{i=1}^n \Omega_i$ where for each $i$, $\Omega_i$ is
an open neighborhood of $\xi_i$ and $\Omega_i \cap \Omega_j =
\emptyset$ if $i \ne j$. Given $\epsilon > 0$, there exists
$\delta > 0$ such that $\max_i |\Omega_i| < \epsilon$ whenever
$|{\cal N}| < \delta$ (an absolute value sign denotes the diameter
in the natural metric).
\end{theorem}

This analysis shows that a differentiable manifold of potentials
which allow for the convergence of cluster expansions is given by
the integral, in the sense of \eqref{integral}, of the set of all
degree $n-1$ polynomials $f$ with all zeros $\xi_i$ distinct, and
such that $f'(\xi_i) \in \Omega_i$ for all $i$, where $\Omega_i$
are nonoverlapping open sets. We denote this manifold by $\JW$.

\subsection{The Fibers of The Vacuum Bundle}
\label{sec:vacbundle}

The following theorem is an analytic statement about the variation
of $\ker(H)$ as we change the base point in the manifold of
coupling constants. As the vectors in $\ker(H)$ are identified
with physical ground states (also called {\it vacua}), Theorem
\ref{thm:grstates-wzmodel}, together with our characterization of
the moduli space $\JW$ of admissible potentials, implies the
existence of a vector bundle built from the vacua, as predicted by
Cecotti and Vafa \cite{Cecotti:1991me}. We propose that results of
this type be termed {\it vacuum bundle estimates}.

\begin{theorem} \label{thm:grstates-wzmodel}
Let $M$ be a complex manifold of dimension $d$, and let $W : M
\times \C \to \C$ be a function which is holomorphic in its
dependence on $m \in M$ and in its dependence on $z \in \C$.
Assume that $W(m,z)$ is polynomial in the $z$ variable with $n =
\deg W$ equal to a constant function on $M$. Assume also that for
each $m \in M$, the $N=2$ Wess-Zumino Hamiltonian $H_W$ defined by
choosing $W(m,z)$ as polynomial superpotential satisfies $n_-(H_W)
= 0$. Let $\V(m)$ denote the ground state subspace of the
Wess-Zumino model defined by $W(m,z)$, i.e. $\V(m) =
\ker(H_{W(m,z)})$. Then $\V$ is a rank $n-1$ holomorphic vector
bundle over $M$.
\end{theorem}

\prooftheorem{thm:grstates-wzmodel} We wish to show holomorphicity
of the ground state vector space; by the vanishing property ($n_-
= 0$), we may restrict our attention to bosonic ground states. We
would like to apply Lemma \ref{lemma:kato-jordan}, but for this we
need holomorphicity of the Hamiltonian.

Our strategy is to first show the desired result for a theory with
an infrared cutoff, and then show that the desired property is
preserved in the limit as the cutoff is removed. Let $\Omega_0$
denote the Fock vacuum. Write $\H_b = \H_{\leq} \otimes \H_{>}$
where $\H_{\leq}$ is spanned by states of the form $R_j \Omega_0$,
with $R_j$ ranging over all finite polynomials in creation
operators $a^*(p)$ for $|p| \, \leq \, (j-1)\kappa$, where
$\kappa$ is some momentum cutoff.

The bosonic, cutoff Hamiltonian for $m \in M$ takes the form
\begin{equation} \label{eq:tensor}
    H_{m,b}(\kappa) = {H_m}^{\leq} \otimes I + I \otimes {H_0}^{>}
\end{equation}
where ${H_0}^{>}$ contains no interacting modes (thus it is
independent of $m$), and ${H_m}^{\leq}$ is unitarily equivalent to
a Schr\"odinger operator $-\tri + V_m$ acting on $L^2(\R^j)$ with
polynomial potential $V_m$.

As $m \in M$ changes holomorphically, it follows from well-known
results of constructive field theory (see for example Arthur
Jaffe's PhD thesis) that the Schr\"odinger operators $-\triangle +
V_m$ meet the conditions of Theorem \ref{thm:schrod}. We conclude
that each of the operators appearing in eq.~\eqref{eq:tensor}
depends holomorphically on the parameters $m$ in theory space.

Since none of our results depend on the global geometry or
topology of $M$, we are free to choose, once and for all, a point
$p\in M$ and a (complex) local coordinate chart $\chi = (z_1,
\ldots, z_N)$ in a neighborhood of $p$. We choose the origin of
the coordinate system so that $\chi = 0$ in $\C^N$ corresponds to
$p \in M$, and prove that the Hamiltonian is holomorphic in $\chi$
at $\chi=0$.

Since the Schr\"odinger operators $-\triangle + V_m$ meet the
conditions of Theorem \ref{thm:schrod}, we infer that
$H_{\chi,b}(\kappa)$ is holomorphic in the complex parameter
$\chi$, in the generalized sense for unbounded operators. This
implies that the cutoff resolvent
\[
    R(\kappa, \chi, \zeta) = (H(\kappa, \chi) - \zeta)^{-1}
\]
is bounded-holomorphic in $\chi$. Jaffe, Weitsman, and Lesniewski
have shown that the cutoff resolvent is norm continuous in
$\kappa$ and moreover
\[
    \lim_{\kappa \to \infty}
    (H(\kappa, \chi) - \zeta)^{-1}
    =
    (H(\chi) - \zeta)^{-1}
\]
We need to show that the norm limit $R(\chi, \zeta)$ is also
bounded-analytic in $\chi$; this will follow if we prove that the
derivative with respect to $\chi$ of the cutoff resolvents
converges, in the limit as the cutoff is removed, to the
derivative of $(H(\chi)-\zeta)^{-1}$.

We have
\begin{equation} \label{eq:deriv-resolvent}
    \frac{\d}{\d \chi}
    (H(\kappa, \chi) + \zeta)^{-1}
    =
    \frac{1}{2\pi i}
    \oint_C (H(\kappa, \chi') + \zeta)^{-1} (\chi' - \chi)^{-2} d\chi'
\end{equation}
where $C$ is a circle in the complex $\chi$-plane around the point
of holomorphicity (in this case $\chi = 0$). The limit of the
derivative of the resolvent as $\kappa \to \infty$ is the limit of
the l.h.s. of \eqref{eq:deriv-resolvent}, which must equal the
limit of the r.h.s. $\ $Since $C$ is compact, the integrand is
uniformly continuous, and hence the $\kappa \to \infty$ limit can
be interchanged with $\oint_C$. Moving the limit inside, we use
the fact that the resolvents $(H(\kappa,\chi) + \zeta)^{-1}$
converge in norm to the resolvent of the limiting theory $(H(\chi)
+ \zeta)^{-1}$. So the limit of the derivative of the resolvent as
$\kappa \to \infty$ is
\[
    \lim_{\kappa \to \infty}
    \left( \frac{\d}{\d \chi} (H(\kappa,\chi) + \zeta)^{-1} \right)
    =
    \frac{1}{2\pi i} \int_C (H(\chi') + \zeta)^{-1}
    (\chi' - \chi)^{-2} d\chi'
\]
which equals the derivative of the resolvent of $H(\chi)$. We
infer by Theorem \ref{thm:unbdd-ops} that the Hamiltonian of the
limiting theory is holomorphic in $\chi$.

The Hamiltonian $H(\chi = 0)$ has a spectral gap above the ground
state eigenvalue.  In fact $H(\chi = 0)$ is essentially
self-adjoint with trace class heat kernel, so the spectrum
consists entirely of isolated points. Therefore Lemma
\ref{lemma:kato-jordan} applies; specifically, we choose the
rectifiable Jordan curve required by the Lemma to be a circle
enclosing only the ground state eigenvalue. In the notation of
Lemma \ref{lemma:kato-jordan}, the vacuum states are basis vectors
for the subspace $M'(\chi)$ and we conclude that $M'(\chi)$  is
holomorphic in a neighborhood of $\chi = 0$. This completes the
proof.

The rank of the vector bundle must be $n_+(H)$, which equals
$\ind(Q)$ by the vanishing theorem. But the latter was shown by
Jaffe \emph{et al} \cite{2DWZW} to be $n-1$.
\endproof


\section{The $tt^*$ Connection}

In this section we present a rigorous construction of a connection
on the vacuum bundle. The connection which we construct was
originally discovered in a physics context by S.~Cecotti and
C.~Vafa \cite{Cecotti:1991me}. This is a generalization to $N=2$
Wess-Zumino field theory of the representation of Berry's
geometrical phase in ordinary quantum mechanics as the holonomy of
a connection on a principal $U(1)$ bundle.

The WZ Hamiltonian in the limit as the cutoff is removed is well
defined on the tensor product $\H = \H_b \otimes \H_f$ (this is
the main theorem of \cite{JLW}). The result holds for a wide
class of superpotentials, thus the fixed Hilbert space $\H$ that
will be necessary to define the $tt^*$ connection exists. The
result on the existence of the vacuum bundle shows that there is
indeed a subspace $V(m)$ in this fixed Hilbert space $\H$ for
each $m$ in the parameter space $\M$ of superpotentials.

A {\it covariant derivative} on a vector bundle $E \to M$ is a
differential operator
\[
\nabla : \Gamma(M,E) \to \Gamma(M, T^*M \otimes E)
\]
satisfying the Leibniz rule: if $s \in \Gamma(M,E)$ and $f \in
C^\infty(M)$ then $\nabla(f \cdot s) = df \otimes s + f \nabla s$.
A covariant derivative so defined automatically extends to give a
map
\[
    \nabla : \Omega^\bullet(M,E) \to \Omega^{\bullet+1}(M,E).
\]

Consider a coordinate chart $U \subset \M$ with local coordinates
$(x^a), \, a=1\ldots n$. Let $V \to \M$ be the vacuum bundle. The
restriction $s|_U$ of a section $s \in \Gamma(\M,V)$ can be
identified via the coordinates $(x^a)$ with a function on $\R^n$
taking values in $\H$, which we denote by $s(x^1, \ldots, x^n)$.
We write $\d_a s$ for the partial derivative $\d s(x^1, \ldots,
x^n) / \d x^a$.

Suppose that the states $| \alpha(x)_i \>, \ i=1, \ldots,
\rank(V)$ form an ON basis of $V(x)$ for each $x \in U$, and vary
smoothly in their dependence on $x$. Equivalently, the
$|\alpha(x)_i \>$ form a local orthonormal frame for $V$.
Consider a curve $\lambda \to x_\lambda$ mapping $(0,1)$ into
$U$. We note that in the difference quotient
\[
    \frac{d}{d\lambda}{\Big |}_{\lambda = 0} |\alpha(x_\lambda)\>
    = \lim_{h \to 0}
    \frac{1}{h} \Big(
    |\alpha(x_{\lambda+h})\> - |\alpha(x_\lambda)\>
    \Big),
\]
$|\alpha(x_{\lambda+h})\>$ and $|\alpha(x_\lambda)\>$ represent
vacuum states of different Hamiltonians, and hence the difference
$|\alpha(x_{\lambda+h})\> - |\alpha(x_\lambda)\>$ is not a ground
state, and even if the spaces $V(x)$ are closed, the partial
derivative $\d_a s$ of a section $s$ can lie outside of $V$.

We define a covariant derivative on $V$ by the equation
\[
(\nabla s)_m \equiv P_{V(m)} (\d_a  s)_m \, dx^a
\]
so that $\nabla s \in \Gamma(M, T^*M \otimes V)$. $P_{V(m)}$ denotes the
projection onto the vacuum subspace $V(m) \subset \H$. A sum over each index
appearing in both upper and lower positions is implied. Thus $\nabla s$ is a
one-form with coefficients in $V$, i.e. a section of
$\Omega^1(M) \otimes V$.

Since the states $|\alpha(x)_j\>$ are locally a basis of $V$, we can
determine the matrix for $\nabla$ in this basis:
\[
\nabla | \alpha_i \> = | \alpha_j \> {\omega^j}_i
\]
where $\omega = ({\omega^j}_i)$ is a matrix-valued one-form.
By the definition of $\nabla$,
\[
P_{V} \d_a  | \alpha_i \> \, dx^a = | \alpha_j \> {\omega^j}_i
\]
Taking the inner product with $\<\alpha_k|$  yields an expression for the
connection forms $\omega_{k i}$
\[
\< \alpha(x)_k | P_{V(x)}\frac{\d}{\d x^a} | \alpha(x)_j \> =  \<
\alpha(x)_k | \alpha_j \> {\omega^j}_i = \omega_{k i}
\]
We now show that for the purposes of computing the connection forms, it is
not necessary to insert the projection operator $P_V$. Since the states
$| \alpha(x)_j \>$ are a local frame for $V$, we can write
\[
P_V = \sum_j | \alpha(x)_j \> \< \alpha(x)_j |
\]
It follows that
\begin{eqnarray*}
\< \alpha(x)_k | P_{V(x)}\frac{\d}{\d x^a} | \alpha(x)_j \> &=& \<
\alpha(x)_k | \sum_j | \alpha(x)_j \> \< \alpha(x)_j |
\frac{\d}{\d x^a} | \alpha(x)_j \> \\
&=& \sum_j \delta_{jk}
\< \alpha(x)_j | \frac{\d}{\d x^a} | \alpha(x)_j \> \\
&=& \< \alpha(x)_k | \frac{\d}{\d x^a} | \alpha(x)_j \>
\end{eqnarray*}

These considerations do not depend in an essential way on the intended
application to $(2,2)$ supersymmetric QFT's. The above discussion in fact
proves the following general existence theorem:

\begin{theorem}
Let $\V : M \to Gr_k(\H)$ be a smooth map from $M$ into the
Grassmannian of $k$-dimensional closed subspaces of a fixed
Hilbert space $\H$. Then under a suitable local condition on the
transition functions, the association $x \to \V(x)$ gives rise to
a $C^\infty$ vector bundle $E \overset{\pi}{\To} M$, where $E =
\bigcup_{x \in M} \V(x)$. This bundle inherits a natural Hermitian
structure $g$ from the Hilbert space inner product, defined by
$g_x(\phi, \psi) = \< \phi | \psi \>$, where $\phi, \psi \in
E_x$. The Levi-civita connection corresponding to this Hermitian
structure is given explicitly by the formula
\[
(\nabla s)_m \equiv P_{V(m)} (\d_a  s)_m \, dx^a \ \ \text{ for }
\ \ s \in \Gamma(E)
\]
In a specific choice of a local orthonormal frame, the connection forms
$\omega_{ki}$ are given by
\[
    \omega_{k i} =
    \Big\< \alpha(x)_k  \Big| \frac{\d}{\d x^a} \Big|
    \alpha(x)_j \Big\> dx^a
\]
\end{theorem}

\subsection{Application: The CFIV Index}

The ground state metric arises in calculations of the CFIV index
\cite{Cecotti:1992qh}, as well as in other important calculations.
The infinite volume theory entails degenerate vacua at +/- spatial
infinity, and what is actually well defined is the trace
$\Tr_{(a,b)}$ over the $(a,b)$ sector, where $a$ and $b$ are
indices which label the different ground states. Physicists
calculate \cite{Cecotti:1992qh} that for a cylinder of length $L$
and radius $\beta$, the CFIV index $Q_{ab}\equiv i\beta L^{-1}
\Tr_{(a,b)} (-1)^F F e^{-\beta H}$ is given by
\begin{equation} \label{cyl}
    Q_{ab}  = -(\beta \, g\d_\beta g^{-1} + n)_{ab}
\end{equation}
where $n$ is the number of fields in the Landau-Ginzburg theory
and $g$ is the ground state metric. Thus the calculation of the
CFIV index in the $(a,b)$ sector is reduced to calculating the
metric $g$. In principle this is done by integrating the $tt^*$
differential equation which $g$ satisfies, however these equations
are complicated. One simplification is to transform to a special
gauge in which the $tt^*$ equation becomes
\be \label{ttstar}
    \bar{\d}_j (g \d_i g^{-1}) = \beta^2 [C_i, g C_j^\dagger g^{-1}]
\ee
where $C_{ij}^k$ is the structural tensor for the chiral ring.

Eq.~(\ref{ttstar}) is an $N \times N$ matrix of differential
equations involving the components of $g$, where $N$ is the number
of ground states, or chiral fields. These equations are
integrable, and in certain cases equivalent to classical equations
of mathematical physics, which are generally Toda systems.
Therefore (\ref{ttstar}) determines the ground state metric
non-perturbatively. Using the resulting solution in \eqref{cyl}
gives the CFIV index. Other $tt^*$ equations include a flatness
condition for the connection, $[D_i, D_j] = 0$ and the
integrability condition for the tensor $C_{ij}^k$, i.e. $D_i
C_{jk}^\ell = D_j C^\ell_{ik}$.

Results of this paper show that the structures (vacuum bundle,
metric $g$) used in the above heuristic argument do exist. Thus
our results are basic for any future rigorous study of the CFIV
invariant in infinite volume.



\section{Holomorphic Quantum Mechanics}
\label{sec:qm}

We describe a model of $N=2$ quantum mechanics with interactions
parameterized by a holomorphic superpotential $W(z)$. The coupling
constant space is usually taken to be $\C^{n+1}$ (a vector in
$\C^n$ corresponds to a coefficient vector for a polynomial $W$ of
degree $n$), although many of the results generalize to the
situation in which we replace $\C^{n+1}$ by an arbitrary Stein
manifold \cite{Klimek:1990jp}. For this reason the model is also
called {\it holomorphic quantum mechanics}.

The Hamiltonian is a mathematically well-defined generalization of
the Hamiltonians of various phenomenological systems. Application
of this model to a system of interacting pions is described in
\cite{JLL}. We prove that the vector space of ground states varies
continuously in the Hilbert Grassmannian, under suitable
perturbations. This is a special case of the fundamental vacuum
bundle estimate which was introduced as Theorem
\ref{thm:grstates-wzmodel}, however the $N=2$ quantum mechanics
model is sufficiently simple that it is possible to understand the
vacuum bundle estimate in an elementary way.

The model we will study is the one-dimensional version of $N = 2$
supersymmetric Landau-Ginzburg quantum field theory. In this
model, $z(t)$ denotes one bosonic degree of freedom, and $\psi_1,
\psi_2$ are fermionic degrees of freedom. The Lagrangian
\[
\L = \abs{\dot z}^2 + i (\bar{\psi_1} \dot \psi_2 +
\bar{\psi_2} \dot\psi_1) + \bar{\psi_1} \psi_1 \d^2 V +
\bar{\psi_2} \psi_2 (\d^2 V)^* - \abs{\d V}^2
\]
is parameterized by $V(z)$, a holomorphic polynomial of degree
$n$ in $z$. In supersymmetric models, the Hamiltonian may be
expressed as the square of a supercharge. The latter is computed
from the supersymmetry transformations and the Noether theorem.
The result of that calculation gives:
\[
H = Q^2 = -\d\bar{\d} - \bar{\psi_1} \psi_1 \d^2 V -
\bar{\psi_2} \psi_2 (\d^2 V)^* + \abs{\d V}^2
\]
This is motivated by the application to a quantum theory with $N =
(2,2)$ supersymmetry, in which we study the space of ground
states:
\[
V = \left\{ \ket{\alpha} \in \H : Q\ket{\alpha} =
Q^\dagger \ket{\alpha} = 0 \right\}
\]
We define a map $\V : \M \to Gr(\H)$, i.e. from the moduli space
$\M$ of admissible supersymmetric quantum theories into the
Hilbert Grassmannian of $\H$, called the {\it vacuum}:
\[
m \overset{\V}{\Mapsto} \ker H(m)
\]

In order to define the vacuum map explicitly, we first review the
results of \cite{JLL}. Every zero mode arises from a pair $(f, g)$
of $L^2(\C)$ functions, where $g$ satisfies the differential
equation
\begin{equation} \label{susy1}
(-\bar{\d}\d + \abs{\d V}^2) g + (\d^2 V / \d V)^* \d g = 0
\end{equation}
and $f$ satisfies the complex conjugate equation. We refer to
\eqref{susy1} as the {\it supercharge-kernel equation}. For $V =
\lambda z^n$, \eqref{susy1} becomes
\begin{equation} \label{susy2}
-\d\bar{\d} g  + (n-1) \bar{z}^{-1} \d g + \abs{ n \lambda z^{n-1} }^2 g = 0
\end{equation}
Representing $z$ in polar coordinates $(r,\theta)$ and writing
$g(r,\theta)$ as a Fourier series in the angular variable
\[
g(r,\theta) = \sum_{m \in \Z} u_m(r) e^{i m \theta}
\]
yields an ODE for the radial functions:
\begin{equation} \label{ode1}
- u_m'' + \frac{2n-3}{r} u_m' +
\left( 4n^2 \lambda^2 r^{2n-2} + \frac{m(m-2n+2)}{r^2} \right) u_m = 0
\end{equation}
This equation takes the general form \eqref{ode2}; we study
regularity of such objects in Lemma \ref{lemma:regularity}.

\begin{lemma}  \label{lemma:regularity}
Solutions of equations of the type
\begin{equation} \label{ode2}
u'' + A r^{-1} u' + \left( B \lambda^2 r^\alpha + C r^{-2} \right) u = 0
\end{equation}
display regularity in the parameter $\lambda$, where
$A, B,$ and $C$ are nonzero real constants.
\end{lemma}

\prooflem{lemma:regularity} A generic second-order initial value
problem of the form (\ref{ode2}) can be transformed into a system
of equations of first order. Such systems are equivalent to vector
integral equations of Volterra type
\begin{equation}
\label{eq:volterra} \b{y}(x; \lambda) = \b{g}(x; \lambda) +
\int_{\alpha(\lambda)}^x \b{k}(x,t,\b{y}(t;\lambda); \lambda) \,
dt .
\end{equation}
Here $x$ and $t$ are always real, but $\b{g}, \b{k},$ and $\b{y}$
may be complex-valued. More than one real or complex parameter is
allowed, i.e. $\lambda \in \R^m$ or $\C^m$. Theorem 13.III in
\cite{Walter} shows that the solution $\b{y}$ to an equation of
the form (\ref{eq:volterra}) is holomorphic in the parameter
$\lambda$.
\endproof

\begin{lemma}
\label{lemma:projectors} Let $f_1, \ldots, f_n$ be continuous maps
from a topological space $\Lambda$ into a Hilbert space $\H$ such
that $V(\lambda) := \span\{ f_1(\lambda), \ldots, f_n(\lambda) \}$
is $n$-dimensional for any $\lambda$. Then $\lambda \Mapsto
V(\lambda)$ is a continuous map into $Gr(\H)$. Moreover, if
$\Lambda$ is a complex manifold and each $f_j$ is holomorphic,
then so is $V(\lambda)$.
\end{lemma}

\prooflem{lemma:projectors} For each $\phi \in \H$, let $N_\phi(A)
= \norm{A \phi}$. The collection $\{N_\phi \mid \phi \in \H\}$ is
a separating family of seminorms on $\B(\H)$, and the associated
topology is the strong operator topology. Now suppose $t \to
\psi(t)$ is a continuous map from $\Lambda$ to the unit ball of
$\H$. Then the projector onto the ray containing $\psi(t)$ is
$P_{\psi(t)} = \ket{\psi(t)} \bra{\psi(t)}$, and
$\norm{P_{\psi(t)}} = \big| \braket{\psi(t)}{\phi} \big|$, which
is continuous in $t$; thus the Lemma is proved for $n=1$. In case
$n = \dim V(t) > 1$, we have $\norm{ P_{V(t)} \phi } \leq
\sum_{i=1}^n \norm{P_{\psi_i}\phi} = \sum_{i=1}^n \abs{
\braket{\psi_i(t)}{\phi} }$, and the desired result follows by an
``$\eps / n$ argument.'' The proof of holomorphicity is similar.
\endproof

Lemma \ref{lemma:regularity} and Lemma \ref{lemma:projectors} together
imply the following
\begin{theorem} \label{maintheorem-qm}
The vector space of vacuum states of the $N = 2$ Landau-Ginzburg
model of quantum mechanics varies holomorphically in the Hilbert
Grassmannian over a moduli space of coupling parameters
diffeomorphic to $\C^n \times (\C -\{0\})$, and determines a
vector bundle of rank $(n-1)$.
\end{theorem}

\prooftheorem{maintheorem-qm} We can write down the zero modes as
explicit functions, and thus there are $n-1$ linearly independent
zero modes if $n = \deg V$. Let $\C[z]_n$ denote the space of
polynomials with complex coefficients of degree exactly $n$. Then
$\C[z]_n$ is the space of $\sum_{k=0}^n a_k z^k$ such that $a_n
\ne 0$, and is therefore isomorphic to the open submanifold $\C^n
\times (\C -\{0\})$ of $\C^{n+1}$. By Lemma
\ref{lemma:regularity}, each of the $n-1$ linearly independent
zero modes is holomorphic as a function of the parameters $(a_0,
\ldots, a_k) \in \C^n \times (\C -\{0\})$. \endproof

\section{Directions for Further Research}

Let the coupling constant space of a family of Wess-Zumino models
be $\M$, and let the vacuum bundle be $\V \to \M$. The ground
state metric $g_{i\jbar}$ is a Hermitian metric on $\V$, and
therefore it defines a geodesic flow on $\M$ in situations when
the vacuum bundle can be identified with the tangent bundle $T\M$.
Renormalization also gives a flow on the moduli space $\M$ of
theories, but in this case there is a preferred vector field $\vec
\beta$ which serves as the dynamical vector field of the flow,
known as the beta function.

In a Euclidean quantum field theory defined by an action $S(g,a) =
\int \sigma(g,a,x) dx$ where $g = (g^1, g^2, \ldots)$ is a set of
coupling constants and $a$ is a UV cutoff, we assume there exists
a one-parameter semi-group $R_t$ of diffeomorphisms on $\M$ such
that the theory $S(R_t g, e^t a)$ is equivalent to the theory
$S(g,a)$ in the sense of correlators being equal at scales $x \gg
e^t a$. The $\beta$ function is defined by $dg^i = \beta^i(g) dt$,
thus the vector field $\vec\beta$ generates the flow.

Zamolodchikov defined a metric $G_{ij}$ on $\M$ which
schematically takes the form
\[
    G_{ij}
    =
    \left. x^4 \big\< \Phi_i(x) \Phi_j(0) \big\>
    \right|_{x^2 = x_0^2}
    \ \text{ where } \
    \Phi_i(x) = \frac{\d}{\d g^i} \sigma(g,a,x)
    \, .
\]
Up to singularities, the flow lines determined by acting on a
single point $g \in \M$ with $R_t$ for all $t \in \R$ coincide
with geodesics of $G_{ij}$.

It would be of fundamental importance to develop a mathematically
rigorous version of the renormalization group for the constructive
Wess-Zumino model considered in this paper, and then in those
cases when the ground state metric $g_{i\jbar}$ computes lengths
of vectors in the tangent bundle $T\M$, to prove an exact
relationship between the ground state metric $g_{i\jbar}$ and
Zamolodchikov's metric $G_{ij}$.

A second important unsolved problem is to determine the largest
possible moduli space for two-dimensional $N=2$ Wess-Zumino
theories in which the vanishing property holds. The cluster
expansion is one of the most refined estimates known for stability
of such theories, and yet the cluster expansion is certainly
weaker than the optimal bound. For these reasons, we expect that
the moduli space we have used in this paper is an open subset of
the optimal moduli space for the vacuum bundle.

A new research direction in functional analysis is suggested
following Theorem \ref{thm:schrod}. Moreover, it is likely that
additional new mathematics would be found in a further exploration
of the interplay between the geometry of the vacuum bundle and the
infinite-dimensional analysis of constructive quantum field
theory.

\section*{Acknowledgements} The author wishes to thank Cumrun Vafa,
Sergei Gukov, Alan Carey, Daniel Jafferis, Xi Yin and especially
Arthur Jaffe for helpful discussions. I am also deeply grateful to
the reviewer for many helpful comments on an early version of this
work.


\end{document}